\documentclass[final,5p,times,twocolumn]{elsarticle}

\usepackage{amssymb}
\usepackage{amsmath}
\usepackage{epstopdf}

\usepackage{subfig}
\usepackage{color}

\graphicspath{{./}{./figures/}}

\newcommand{\colorme}[1]{#1}

\begin{document}

\begin{frontmatter}



\title{%
Application of Tomographic Reconstruction Techniques for\\Density Analysis of Green Bodies
}

\author{M. Scot Swan}
\ead{matthewscot.swan@unitn.it}
\author{Andrea Piccolroaz}
\ead{andrea.piccolroaz@unitn.it}
\author{Davide Bigoni\fnref{fn1}}
\ead{bigoni@ing.unitn.it}

\address{Dipartimento di Ingegneria Civile, Ambientale e Meccanica \\
Universit\`a di Trento \\ Via Mesiano 77, I-38050 Trento, Italia}

\fntext[fn1]{Corresponding author.}



\begin{abstract}
Progress in the manufacturing of ceramics, but also of sintered metals, strongly relies on the evaluation of the density distribution in green bodies.
This evaluation is crucial from many points of view, including the calibration of constitutive models for \textit{in-silico} simulation of densification processes.
To this end, X-ray tomography and other techniques are possible but can be unmanageable for some institutions.
Therefore, a destructive method is introduced in the present article to measure the density field of a green body sample using a CNC mill, an analytical balance, and analysis techniques from the field of computational tomography.
A virtual experiment is presented where the method is used to reconstruct a simulated green body density field and is found to satisfactorily correspond to the original solution.
The green body density field of a truncated cylinder made of alumina powder is evaluated using this method and the reconstructed field is presented.
\end{abstract}


\end{frontmatter}


\section{Introduction}
\label{sec:introduction}

The ceramics industry is interested in increasing efficiency, reducing waste, and, therefore, reducing costs.
During the production process the extent of heterogeneity in the green body directly influences the final geometry, strength, and hardness after sintering\cite{Briscoe1998}.
Not only do these variations usually decrease product performance, they also amplify uncertainty in material behavior which is unacceptable when producing high-performance ceramics.
Currently, the ceramics industry heavily relies on the process of trial-and-error to determine optimal mold geometry and forming pressures for a given piece\cite{Carlone2006}.

High-performance ceramics are used in many sectors and are subjected to many different types of environments.
Some usage examples include: refractory products subject to extreme temperatures, piezoelectrics subject to extreme loading or electric fields, or ceramic plates subject to shock loading.
Each of these use-cases requires predictable performance which is often limited by the uncertainty in macroscopic mechanical behavior of the piece.

Density inhomogeneities in green bodies are associated with stress variations which are usually caused by defects in the production process.
However, when final residual stress fields can be predicted and utilized in the design process, these residual stresses can be used to pre-stress the sample to make it more resilient for the intended use-case.

Because so much of the final performance of the ceramic piece is dependent on the density of the green body, many different techniques have been developed to measure internal density.
The most simple method for measuring bulk density accurately is to use Archimedes' principle with mercury displacement instead of water.
More technical methods put inclusions in the powder before compaction in a known configuration, position, or concentration and infer the final density from the final positions of the inclusions.
Some inclusions that have been used are: layers of colored powder\cite{Briscoe1998}, layers of film, or a thin lattice made of lead\cite{Kamm1948}.
A more recent method for density measurement has been \colorme{to} measure x-ray attenuation to measure the average bulk density along the path of an x-ray\cite{Amoros2010}.
The main benefits of this last method are that it is non-destructive, rapid, and can be used on green bodies as well as sintered pieces.

However, the most commonly used density measurement technique in use in the literature today is to utilize surface hardness measures, either from indentation or from scratching, and convert them to density measures using a table that correlates hardness to density\cite{Fleck1992}\cite{Rajab1985}\cite{Briscoe1997}.
The \colorme{three} primary drawbacks to this method are that the table correlating density to hardness must be produced (either by experimentation or making material assumptions)\colorme{, it is a destructive method if more than surface density is desired,} and that the sample must have sufficient cohesion to be worked, scratched, or indented without failure. 
The latter drawback is usually overcome by partially sintering the green body before analysis.

Finally, it is worth mentioning that even the method of X-ray absorption requires a correlation table and calibration to set the relationship between the gray level of the X-ray images and the bulk density\cite{Amoros2010}.

The density evaluation method introduced in the present article is intended to overcome the hurdles of high cost, adding inclusions to the sample, steep learning curves to perform or analyze the measurement, and the need for previously-developed calibration tables.
By decreasing the requirements to make these measurements, small laboratories or universities with limited equipment and budgets can perform 3D density measurements.
As it makes use of simple computer numerical control (CNC) mills and an analytical balance, the accuracy of the analysis is directly related to the accuracy of the equipment and, to a greater extent, the number of data points taken for the measurement.
\colorme{Therefore, the method can be adapted to the accuracy needs of each situation.}

\section{Presentation of the Method}
\label{sec:presentationofthemethod}

The present method for density distribution evaluation is defined in the following steps:
(i) a CNC mill is used to incrementally remove mass from a green body in parallel strips and an analytical balance is used to weigh the sample before and after each strip to obtain the corresponding lost mass;
(ii) a collection of strips for a given transverse section are combined to make a single projection;
(iii) steps (i) and (ii) are repeated in a different direction at least one additional time (see note below);
(iv) these projections and the known geometry (defined by the path, milling bit used, and depth of the CNC mill) of the piece can be used in a tomographic reconstruction routine;
(v) repeat steps (i) through (iv) for each transverse slice in the sample to create a 3D reconstruction.

Because the present technique requires at least two projections to reconstruct the density field and the projection method is a destructive method, it is advisable to either produce multiple samples (one per projection) or utilize symmetry of the body to get multiple projections from a single sample.
Multiple projections from the same sample can be accomplished by milling for one projection on one symmetry section and changing directions for the other symmetry sections.

\begin{figure}
\centering
\includegraphics[width=0.80\columnwidth]{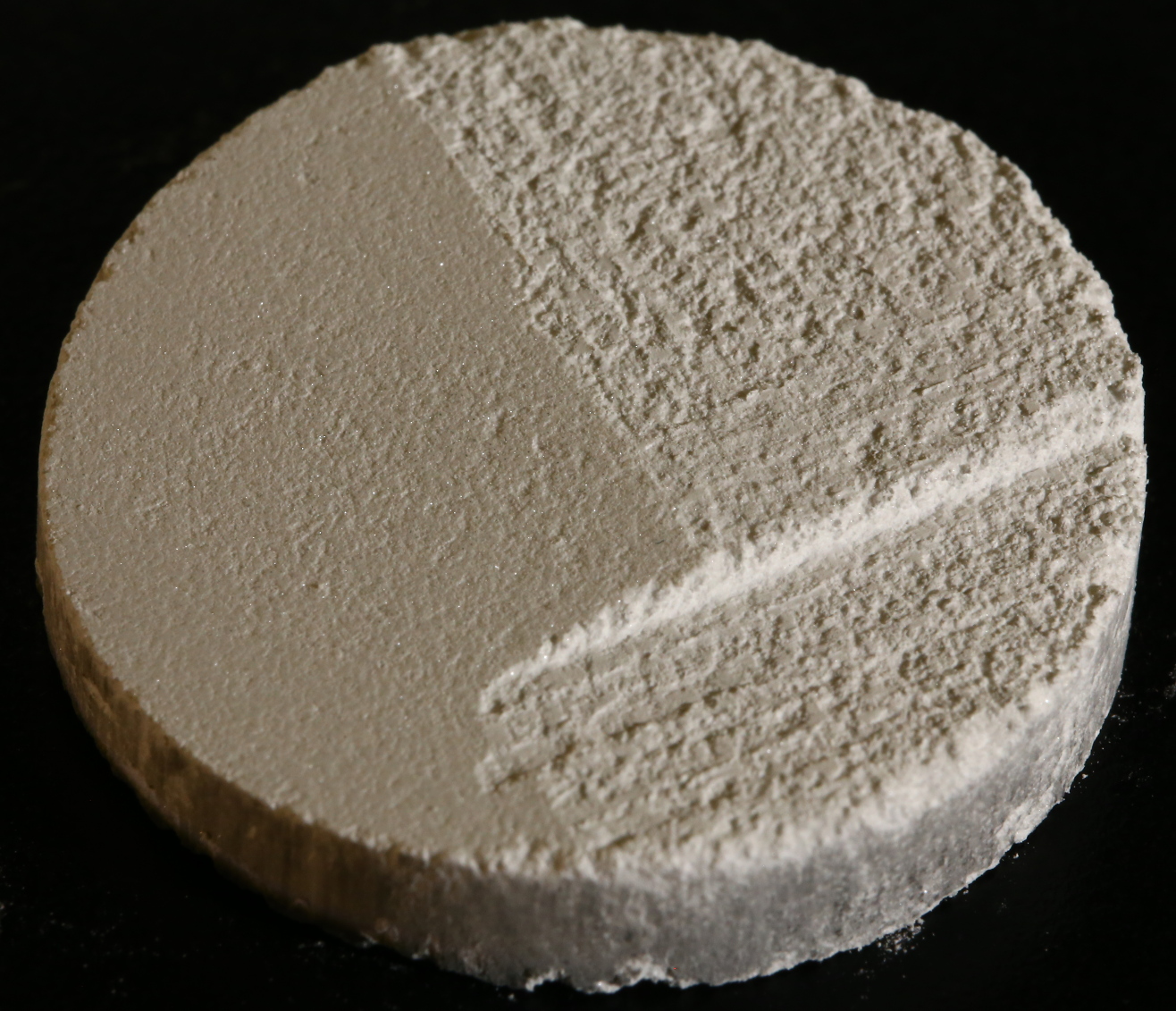}
\caption{Texture of a $10\textrm{g}$ alumina green body formed by $120\textrm{MPa}$ mean axial stress in the shape of a truncated cylinder with a $30\textrm{mm}$ diameter and a $10^\circ$ incline in the process of being milled. The left portion of the sample is the smooth, inclined surface that was created by the mold during pressing. The other surfaces are created by the milling process and demonstrate the rough but uniform texture produced by the CNC mill in the cutting process. The vertical difference between successive layers is $1\textrm{mm}$ (e.g. the step between the surfaces in the top-right and bottom-right of the image).}
\label{fig:texture}
\end{figure}

The basic concept for machining ceramics while in the green body state is discussed by Su \textit{et al.}\cite{Su2008}, but here it is applied to density evaluation, not shaping, finishing, or rapid prototyping.
For a green body of compressed alumina powder, a CNC mill proved to be able to easily mill the body with sufficient precision.
\colorme{The largest source of precision loss in the milling stage was found to be grains becoming dislodged and subsequently lost, leaving a pocked surface instead of a smooth surface.}
This was found to be invariant of the milling speed or the rotational velocity of the bit.
The milled mass needs to be removed after milling each strip, which can be accomplished with a vacuum or compressed air to either suck up or blow away the filings.
For all but the smallest sections, the magnitude of this effect was not excessive and did not invalidate the measurements.
A sample of pressed alumina powder that is in the process of being milled is presented in Figure \ref{fig:texture}.

\section{Tomographic Reconstruction}
\label{sec:tomographicreconstruction}

The mathematical basis of Tomography was laid by Johann Radon in 1917 with his seminal paper that proved that a 2D density function can be exactly reproduced from an infinite number of 1D projections\cite{Radon1986}.
In this paper, he describes what later became known as the Radon transform (the projection step), the output of which is a sinogram, and the inverse Radon transform (the reconstruction step).
These processes are still the basis for most reconstruction techniques.

The birth of modern computer tomography occurred around 1970 when computers with sufficient memory and processing power became available to researchers.
One of the first iterative computational algorithms for reconstructing data from projections was published by Gordon, Bender, and Herman\cite{Gordon1970}.
Their paper introduces an iterative method for solving a system of ill-constrained equations to produce a useful image that approximates the original structure.
While there are currently many different algorithms in use, some more recent reconstruction methods have been reported to be able to reproduce internal structures exactly from highly incomplete frequency information\cite{Candes2006}.

\subsection{Theoretical basis of tomography}
\label{subsec:theory}

\begin{figure}
\centering
\includegraphics[width=0.7\columnwidth]{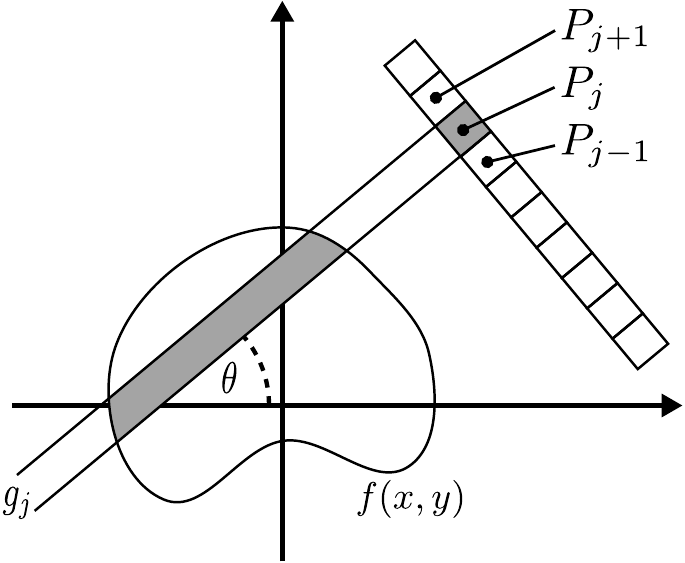}
\caption{A body with a continuously-varying density field $f(x,y)$ is projected in the direction $\theta$ and is gathered into $m$ bins denoted $P_j$ of the projection $P$. Multiple projections along different values of $\theta$ are used to reconstruct the density field $f(x,y)$.}
\label{fig:projectionexample}
\end{figure}

The inverse Radon transform is simply solving a (typically non-linear and ill-constrained) system of equations that are derived from a set of projections.
Each equation in the system represents an element of a single projection (see Figure \ref{fig:projectionexample}).
The formulation is written with a 2D density field $f(x,y) \in [0, \infty)$, a projection angle $\theta \in [0^\circ, 180^\circ)$, and a projection element $P_j$ associated with a projection strip $g_j$ such that
\begin{equation}\label{eqn:integralprojection}
P_j = \int\limits_{g_j} f(x,y) dydx,
\end{equation}
where $j = 1,...,m$.
The projection $P$ is the set of projection elements for one projection direction $\theta$.
For a \colorme{more accurate} reconstruction, \colorme{more} projections with different projection directions $\theta_1, \theta_2 ... \theta_m$ are required.

\begin{figure}
\centering
\subfloat[Phantom]{\label{fig:sinogramexamplea}\includegraphics[width=0.5\columnwidth]{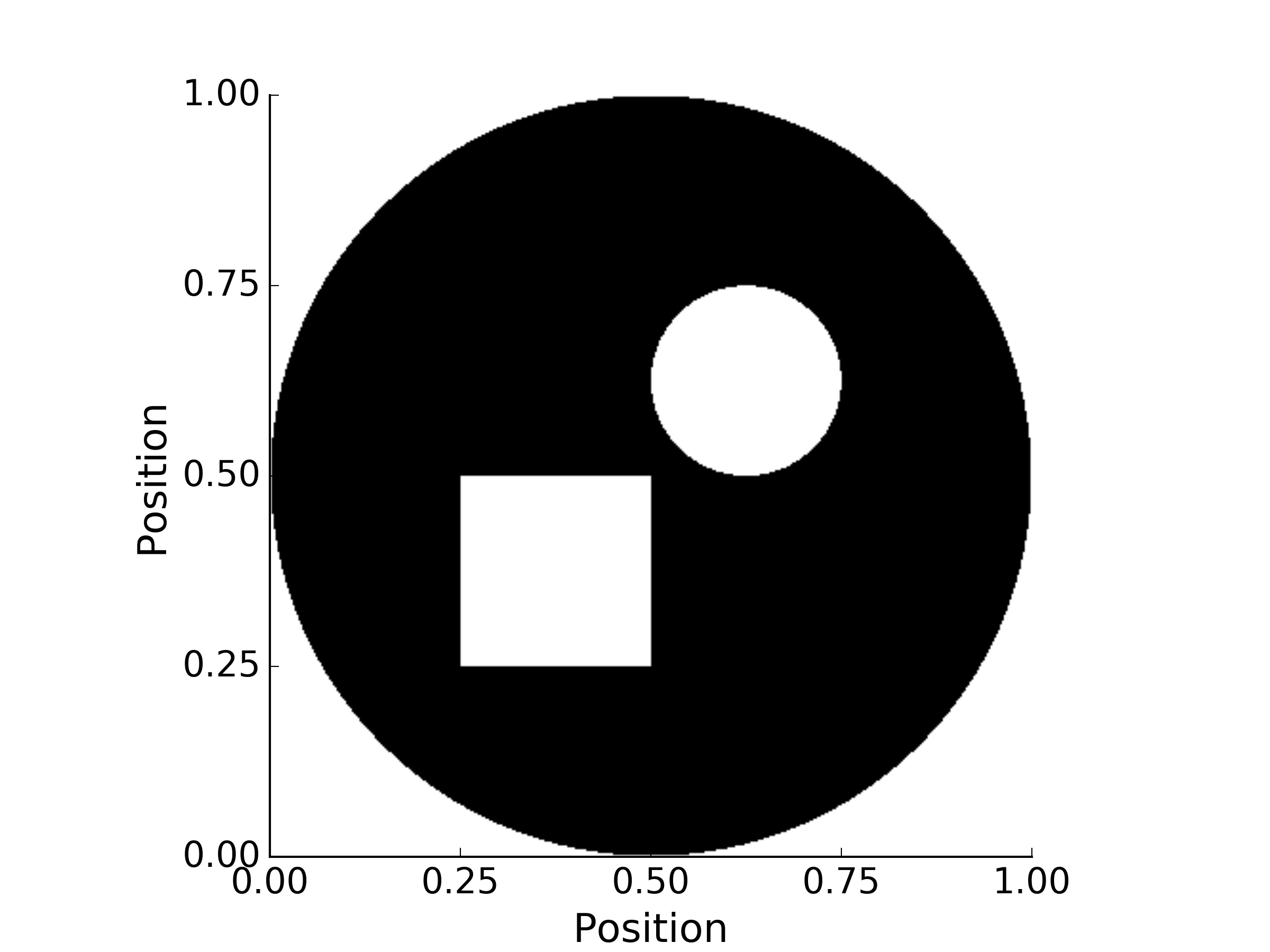}}
\subfloat[Sinogram]{\label{fig:sinogramexampleb}\includegraphics[width=0.5\columnwidth]{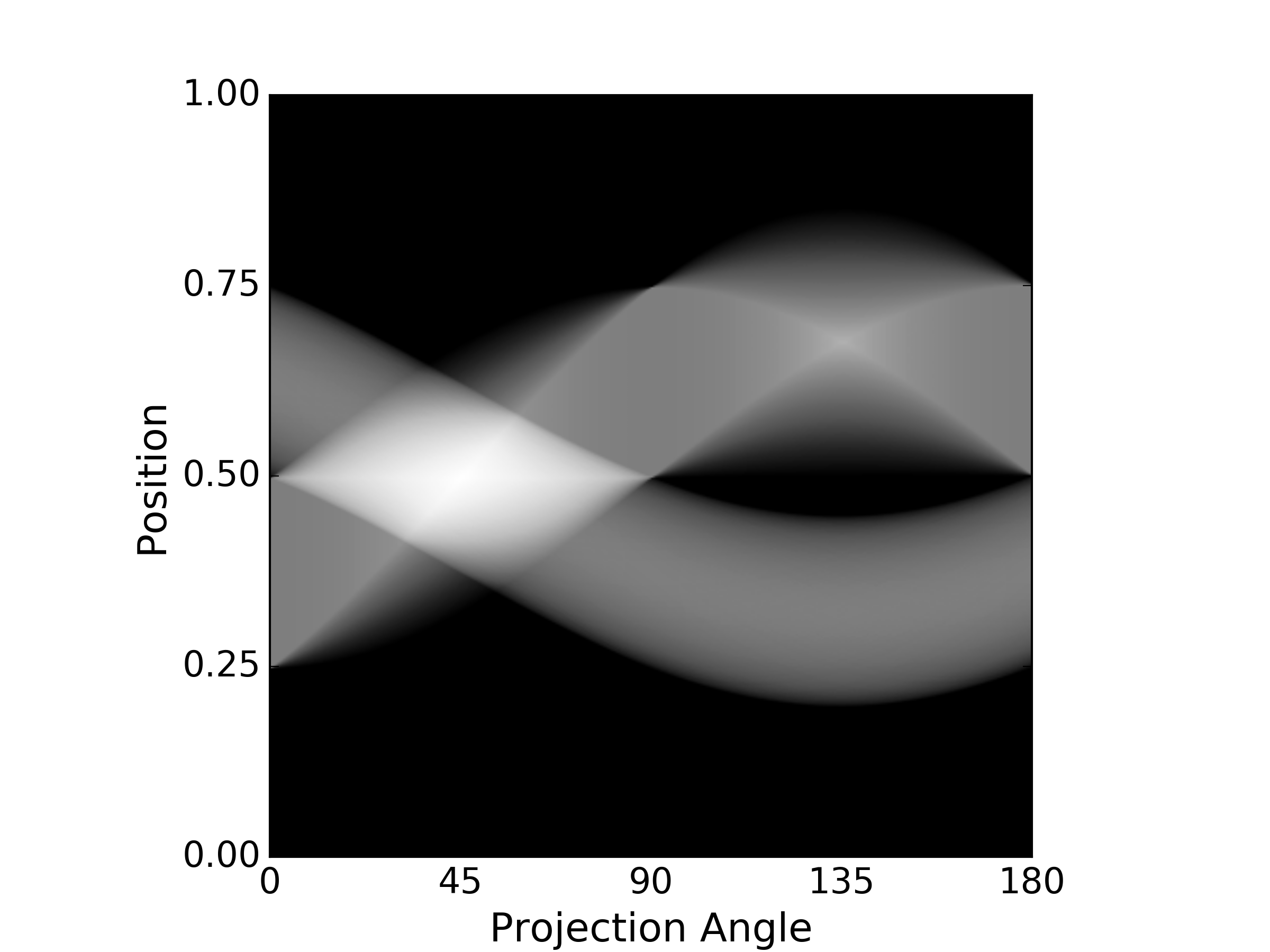}}
\caption{An example of a phantom/sinogram pair. The sinogram was produced using the open-source \texttt{scikit-image} Python library. The analysis domain consists of the inscribed black circle with the white circle and square inclusions. The analysis domain is circular such that the cross-section width is uniform for all projection angles. Notice that, as the projection direction changes in the sinogram, the circle's projection is constant while the square's projection has two intertwining density peaks (from the corners of the square) that both have a period of $180^\circ$.}
\label{fig:sinogramexample}
\end{figure}

In current x-ray tomography, hundreds of \colorme{high-resolution} projections can be taken for a single reconstruction.
These projections are usually compiled into a sinogram that can succinctly convey complete projection information in all directions in one plot (see Figure \ref{fig:sinogramexample}), but is not generally human-readable.
The sinogram is constructed by representing each projection as a column of an image in a sequential manner according to the projection angle.
However, when full projection data are not available, the sinogram is only defined for specific angle values.
There are two methods to handle incomplete projection data: attempt to interpolate the sinogram\cite{Kostler2006} or perform the reconstruction using only the measured projections\cite{Gordon1974}.
This paper follows the latter method as most research involving interpolating sinogram data does not attempt to interpolate over breaks larger than 30 degrees.
Three of the most common types of reconstruction methods are filtered back projection, algebraic reconstruction technique (ART), and 2D Fourier reconstruction\cite{Bruyant2002}.

\subsection{Algebraic Reconstruction Technique}

The algebraic reconstruction technique (ART), which is the reconstruction technique used in this paper, is one of the first iterative reconstruction methods that was developed in 1974 by Gordon and Herman\cite{Gordon1974}.

This method updates the $i$th reconstructed discretized field value at the $q$th iteration $f^q_i$ by enforcing the reconstruction projection element $P^q_j$ to agree exactly with the measured projection element $P_j$.
This is accomplished by evaluating the difference between the measured projection element's value and the value of the current iteration and calculating a correction factor.
If the correction factor is applied additively the method is called the additive ART or, if applied multiplicatively, it is called multiplicative ART.
The additive and multiplicative ART iterative formulas are, respectively,
\begin{equation}\label{eqn:additiveart}
f^{q+1}_i = f^{q}_i + \frac{P_j - P^q_j}{N_j} f^q_i
\hspace{10pt}
\forall f_i \in g_j,
\end{equation}
and
\begin{equation}\label{eqn:multiplicativeart}
f^{q+1}_i = \left(\frac{P_j}{P^q_j}\right) f^q_i
\hspace{10pt}
\forall f_i \in g_j,
\end{equation}
with
\begin{equation}\label{eqn:initialization}
f^0_i = \hat{f}  = \frac{\sum\limits_j P_j}{n}
\hspace{10pt}
\forall f_i,
\end{equation}
where $N_j$ is the number of discretized field elements in the projection strip $g_j$ and $n$ is the total number of discrete values comprising the reconstructed field.
Because the total slice mass is independent of the projection direction, it does not matter which projection is used for initializing the reconstruction field.
Equations (\ref{eqn:additiveart}) or (\ref{eqn:multiplicativeart}) are applied iteratively for each element in each projection until an equilibrium condition is met.

As there is always noise in data acquisition (both epistemic and aleatory uncertainty), it is not uncommon that all of the projection constraints are not able to be exactly met simultaneously.
In an attempt to overcome this limitation, many different convergence criteria have been suggested over the years, including in Gordon's original publication.
The three primary convergence criteria are based on the discrepancy $D$, entropy $S$, and variance $V$ of the reconstruction.
These are defined as
\begin{equation}
D = \sqrt{\frac{1}{m}\sum\limits^m_{j=1}\frac{P_j-P^q_j}{N_j}},
\end{equation}
\begin{equation}
S = \frac{-1}{\ln n} \sum\limits^n_{i=1} \left(\frac{f^q_i}{\bar{f}}\right) \ln\left(\frac{f^q_i}{\bar{f}}\right),
\end{equation}
and
\begin{equation}
V = \sum\limits^n_{i=1}(f^q_i-\bar{f})^2,
\end{equation}
where $\bar{f}$ is the arithmetic mean of the reconstructed field.
As the iteration number increases, $D$ approaches zero and both $S$ and $V$ tend to a minimum.
If the discrepancy does not converge to zero, the reconstruction can be considered converged when the changes in $S$ and/or $V$ are sufficiently small.
There are methods to alleviate some of the problems associated with non-convergence, for example, by interleaving iterations of additive and multiplicative ART or by applying relaxation factors to the correction factor for additive ART\cite{Kak1988}.
In this article, the interleaved method was implemented to improve convergence.

The fidelity of the reconstruction can be enhanced by applying constraints to the reconstruction algorithm.
An obvious constraint for reconstructing density fields is to require each element in the reconstructed field to be non-negative during every iteration or to only reconstruct the field over a specified domain.
The latter can be accomplished simply by setting $f^0_i = 0$ for all field elements outside of the reconstruction domain at each iteration.

\subsection{Number of Projections}
\label{subsec:numberofprojections}

The real benefit of using tomographic techniques to reconstruct a density field inside a ceramic sample is that it increases the resolution of the field while decreasing the amount of labor otherwise required to measure those values.
Obviously, a researcher could cut the sample into an arbitrary number of sections and measure and weigh them to determine their density.
However, for a square domain spanned by $n$ elements in each direction, this would require $n^2$ measurements whereas reconstructing the field using $p$ projections of $n$ elements each, the total number of measurements is $np$.
When less precise data is acceptable, \colorme{such as for initial numerical model validation (the authors' target use-case),} the minimum $2n$ measurements can provide satisfactory aggregate information over the entire green body slice - \colorme{which is} much more efficient than the original $n^2$ measurements.

\subsection{Strengths and weaknesses of 2-projection ART}
\begin{figure*}
\centering
\subfloat[Image]{\label{fig:og}
  \includegraphics[width=0.24\textwidth]{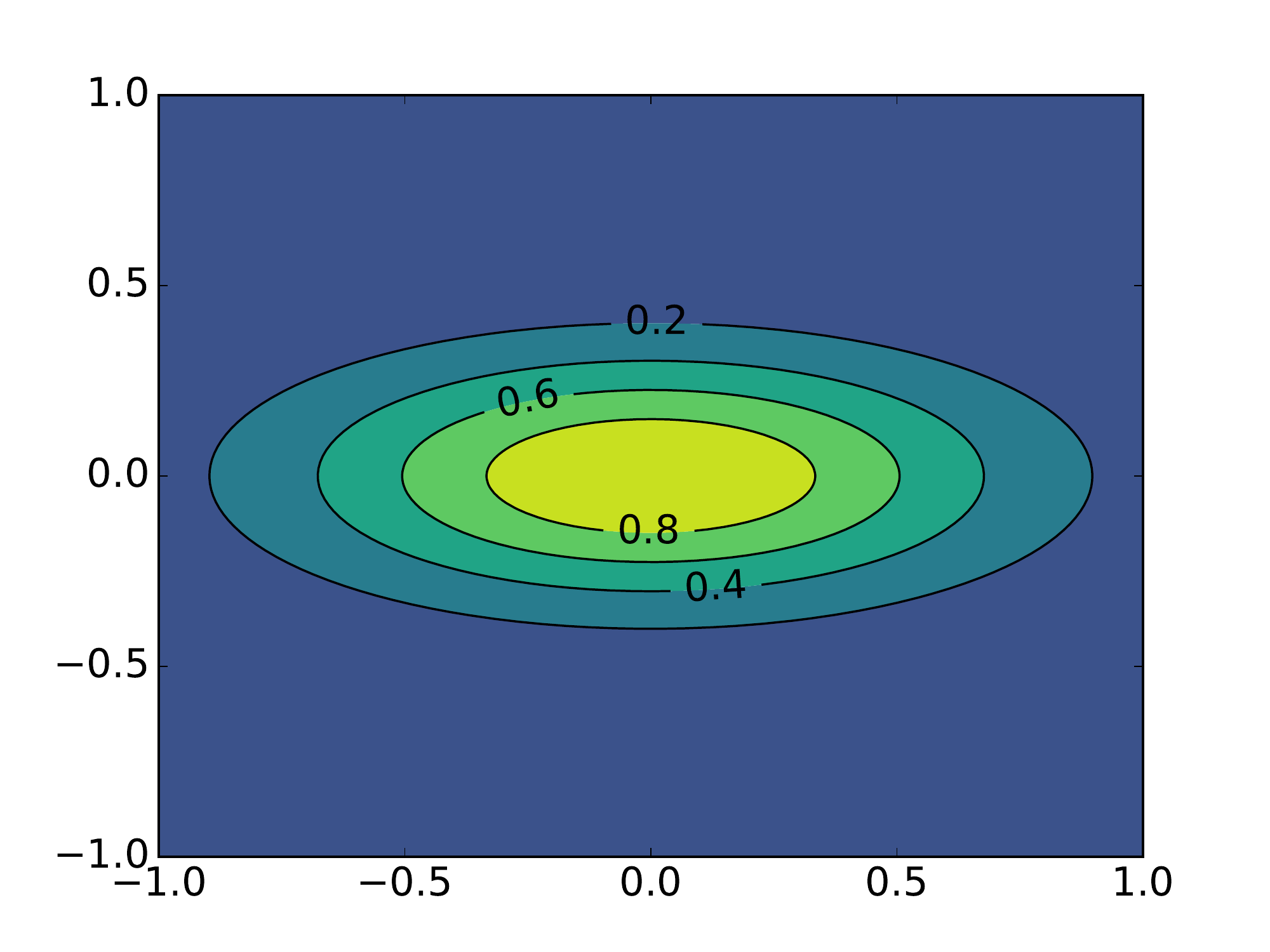}}
\subfloat[Reconstruction]{\label{fig:rog}
  \includegraphics[width=0.24\textwidth]{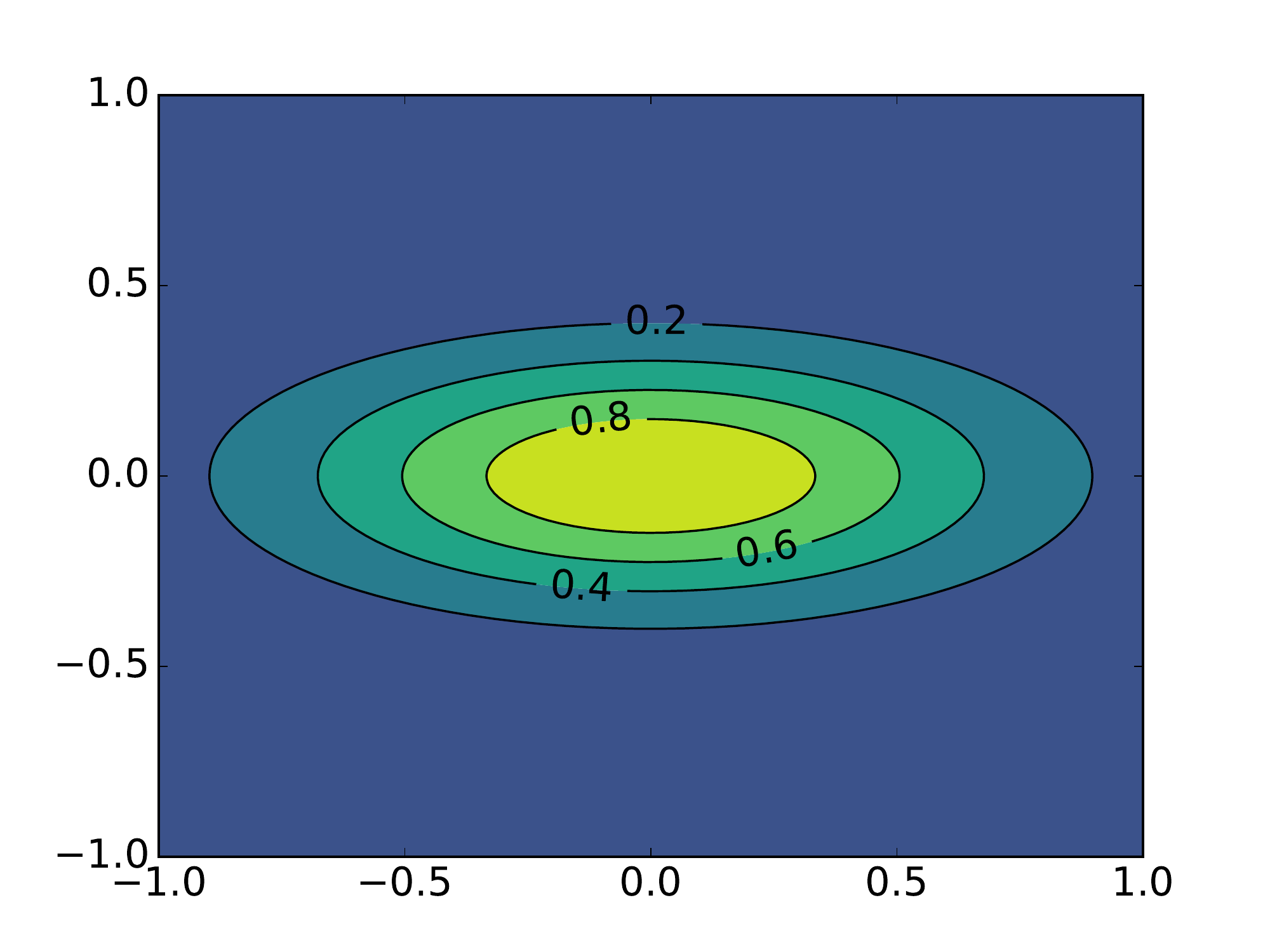}}
\subfloat[Rotated Image]{\label{fig:org}
  \includegraphics[width=0.24\textwidth]{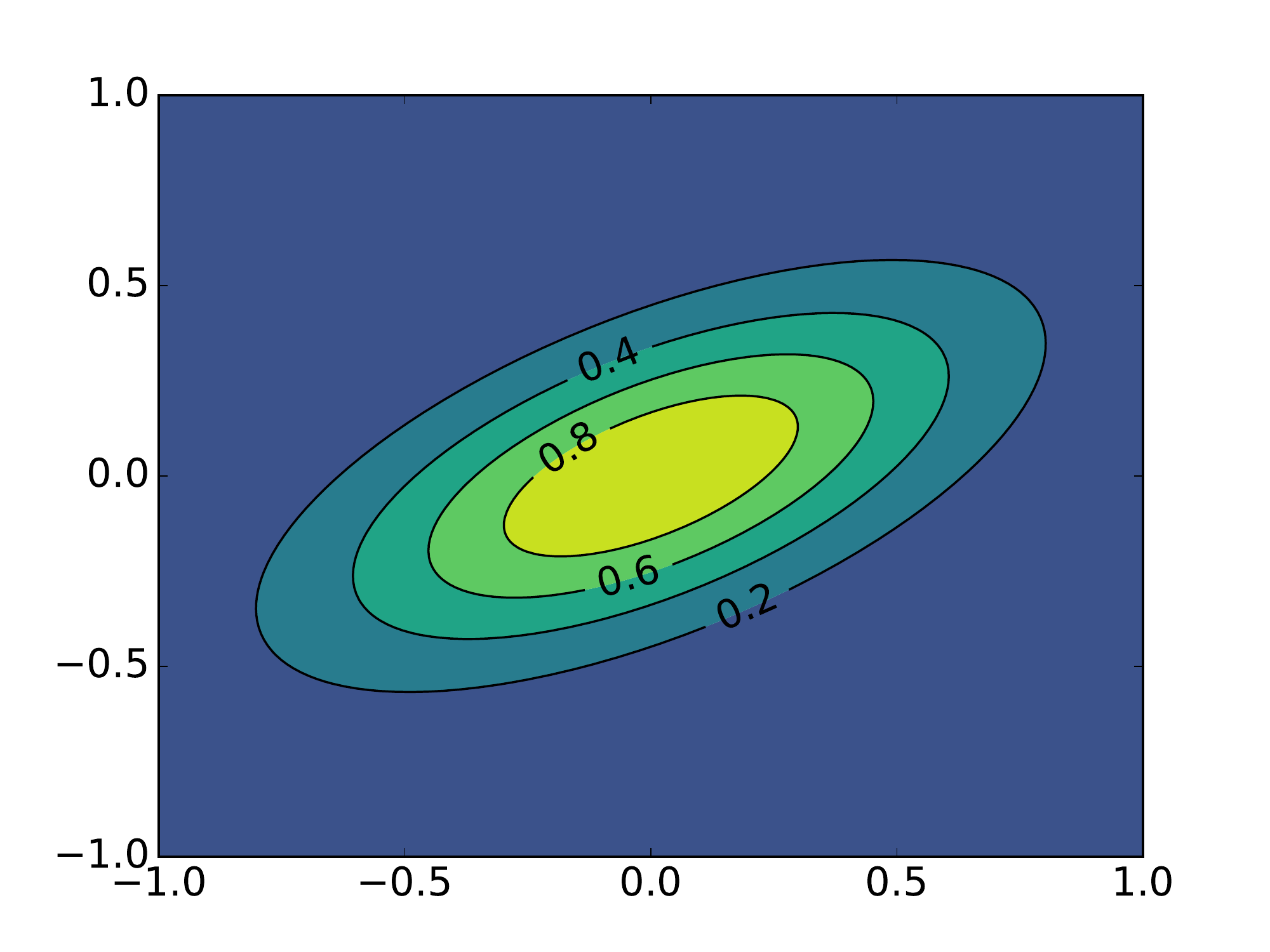}}
\subfloat[Rotated Reconstruction]{\label{fig:rorg}
  \includegraphics[width=0.25\textwidth]{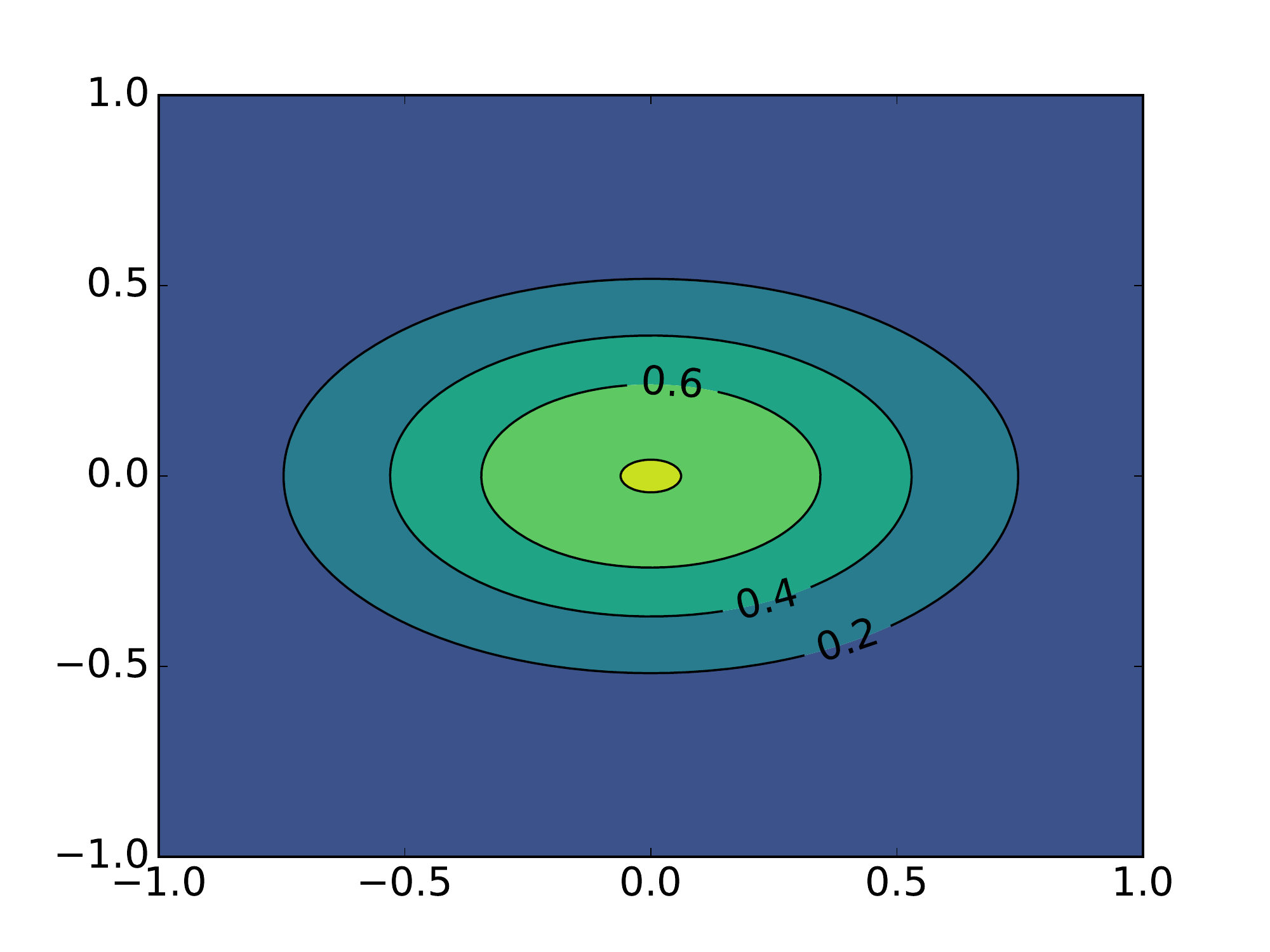}}
\\
\subfloat[Image]{\label{fig:iog}
  \includegraphics[width=0.24\textwidth]{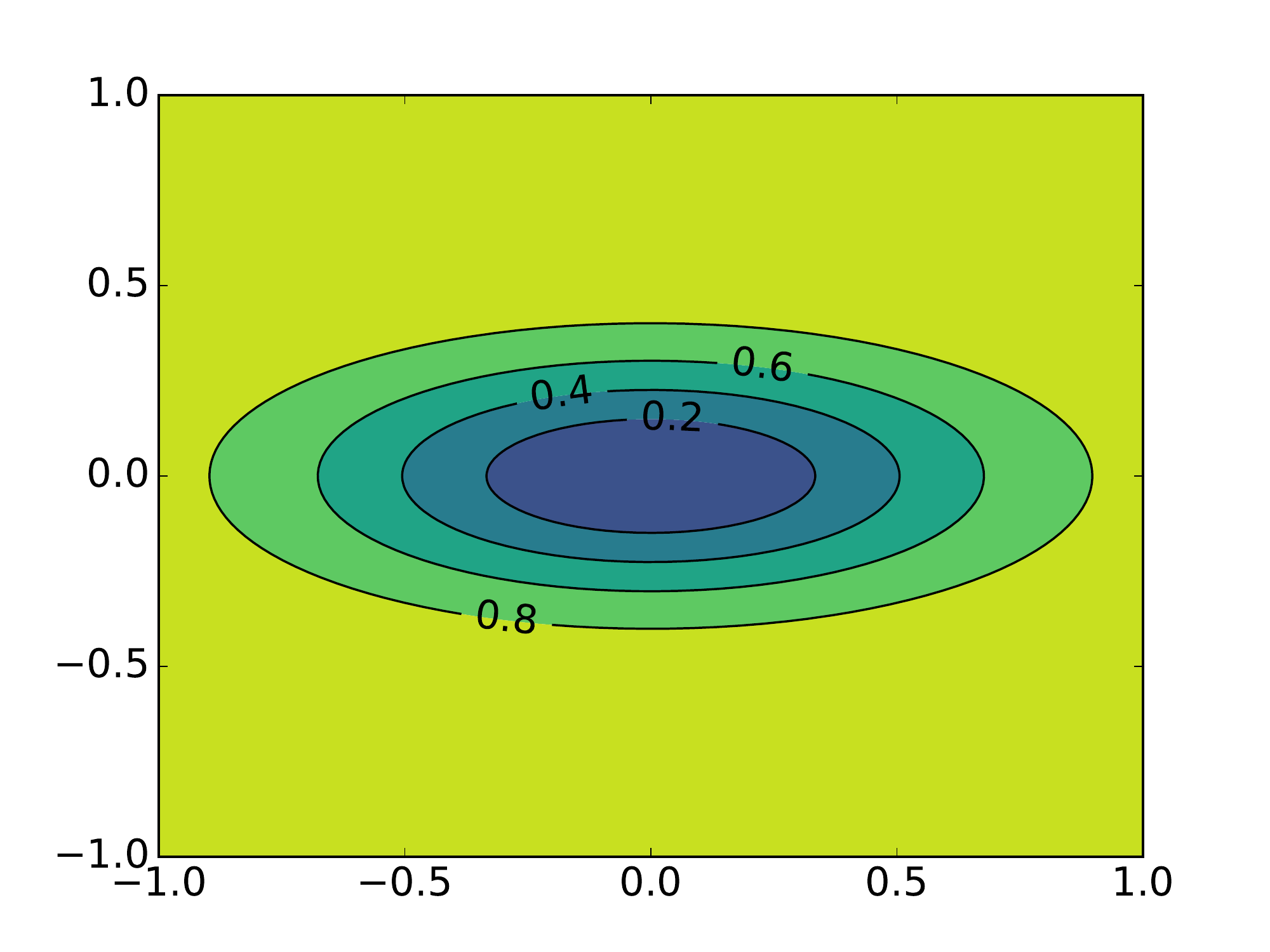}}
\subfloat[Reconstruction]{\label{fig:riog}
  \includegraphics[width=0.24\textwidth]{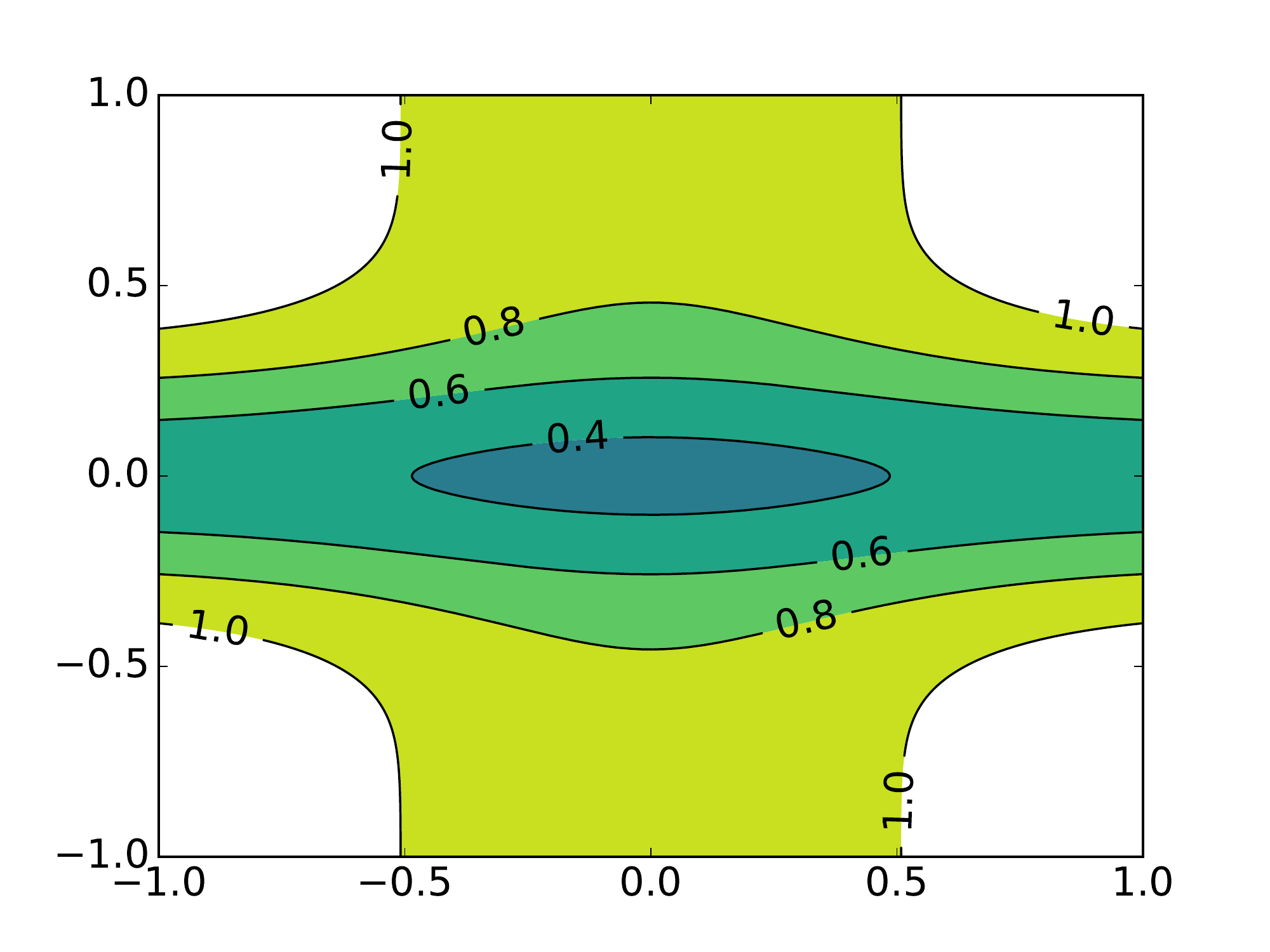}}
\subfloat[Rotated Image]{\label{fig:iorg}
  \includegraphics[width=0.24\textwidth]{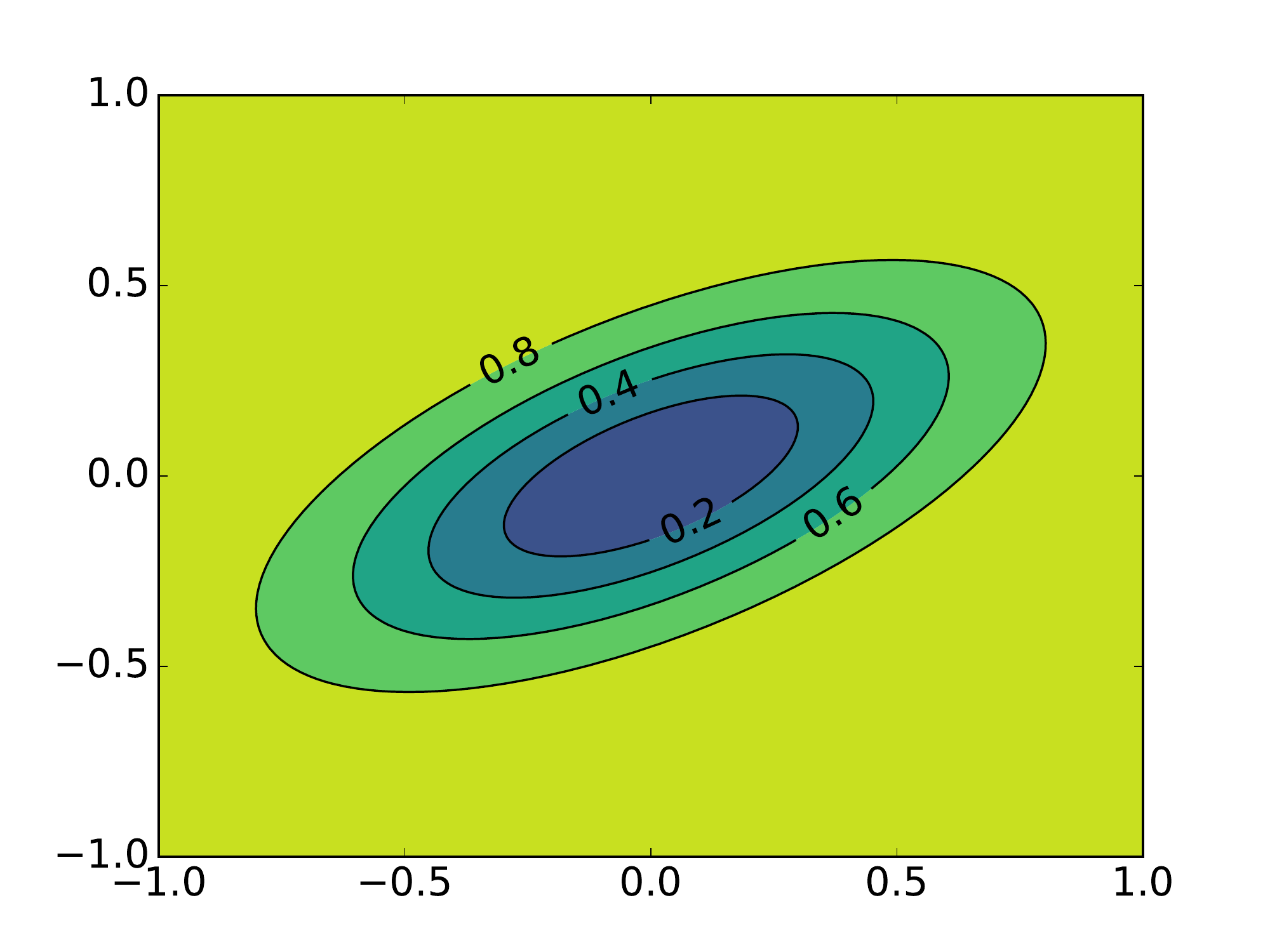}}
\subfloat[Rotated Reconstruction]{\label{fig:riorg}
  \includegraphics[width=0.24\textwidth]{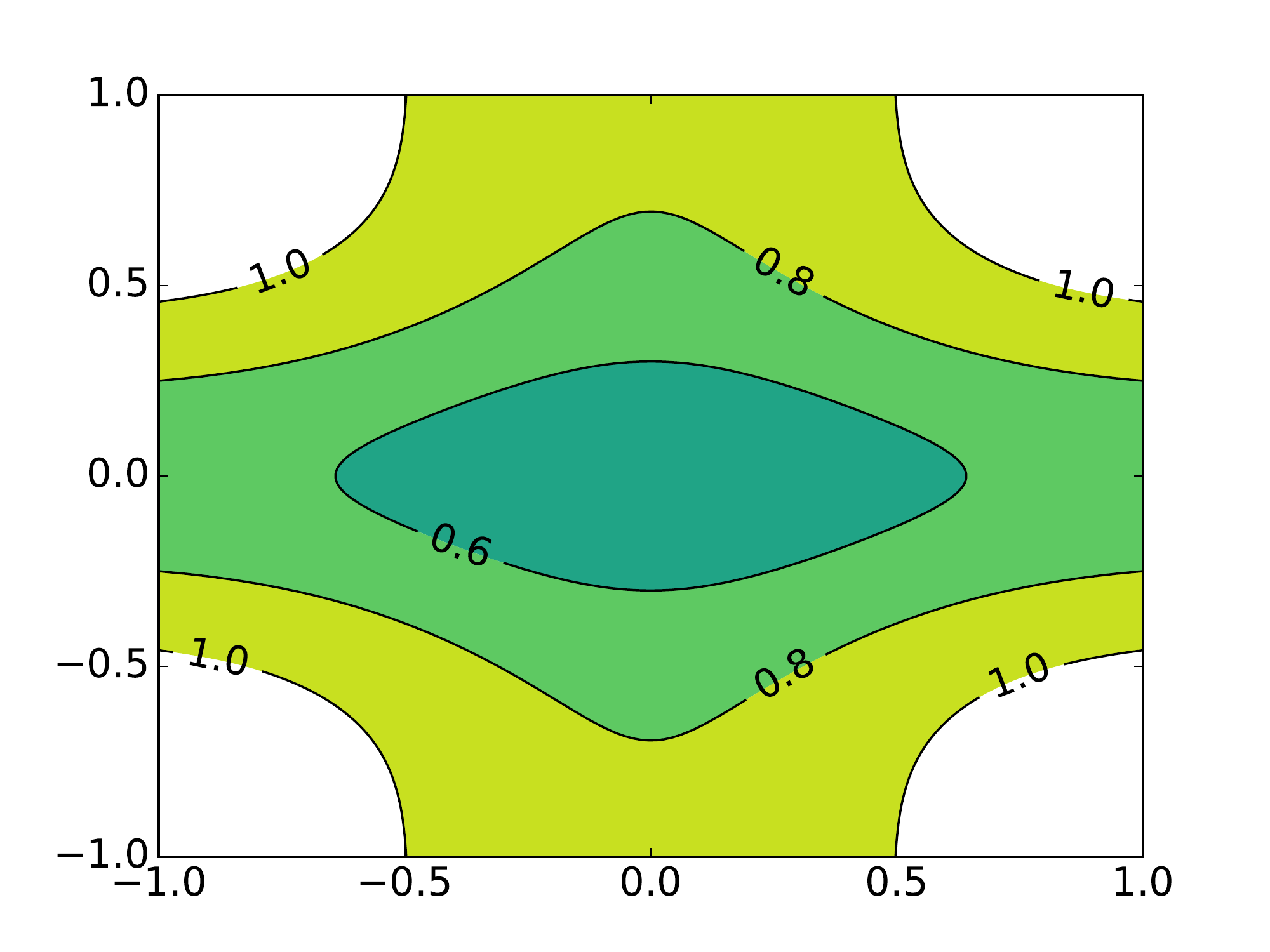}}
\caption{%
A demonstration of the strengths and weaknesses of the 2-projection (vertical and horizontal) multiplicative algebraic reconstruction technique (ART).
The plots in the upper row are based on the function $f(x,y) = e^{-(2x^2+10y^2)}$ with the bottom row based on the function $g(x,y) = 1.0 - f(x,y)$.
As is evidenced by \protect\subref{fig:rorg} and \colorme{\protect\subref{fig:riorg}}, the 2-projection method cannot capture the superimposed rotations present in \protect\subref{fig:org} and \protect\subref{fig:iorg}, respectively, as the principal axes of the function and the projection directions do not coincide.
Note: the reconstruction in \protect\subref{fig:rog} is an exact reconstruction of the image in \protect\subref{fig:og}.%
}
\label{fig:strengthsandweaknesses}
\end{figure*}

Of course, the greater the number of projections the better the accuracy of the reconstructed field.
One of the greatest drawbacks to only using 2 projections is the inability to detect field characteristics at an inclined angle with respect to the projection directions\cite{Raparia1997}.
For this purpose, this method is best used when some \textit{a priori} knowledge can be applied to the analysis such that the projection directions are in line with the natural orientation of the reconstructed field.
This method has been found to be able to exactly reproduce unimodal density fields when principal axes are aligned with the projection directions (see Figure \ref{fig:strengthsandweaknesses}).
If \textit{a priori} knowledge of the field is not available or the field has many local effects, additional projections should be incorporated to reconstruct the structure with greater accuracy.

\section{Virtual Experiment}
\label{sec:virtualexperiment}

\begin{figure}
\centering
\subfloat[Simulation]{\label{fig:virtexpsimulation}
  \includegraphics[width=0.49\columnwidth]{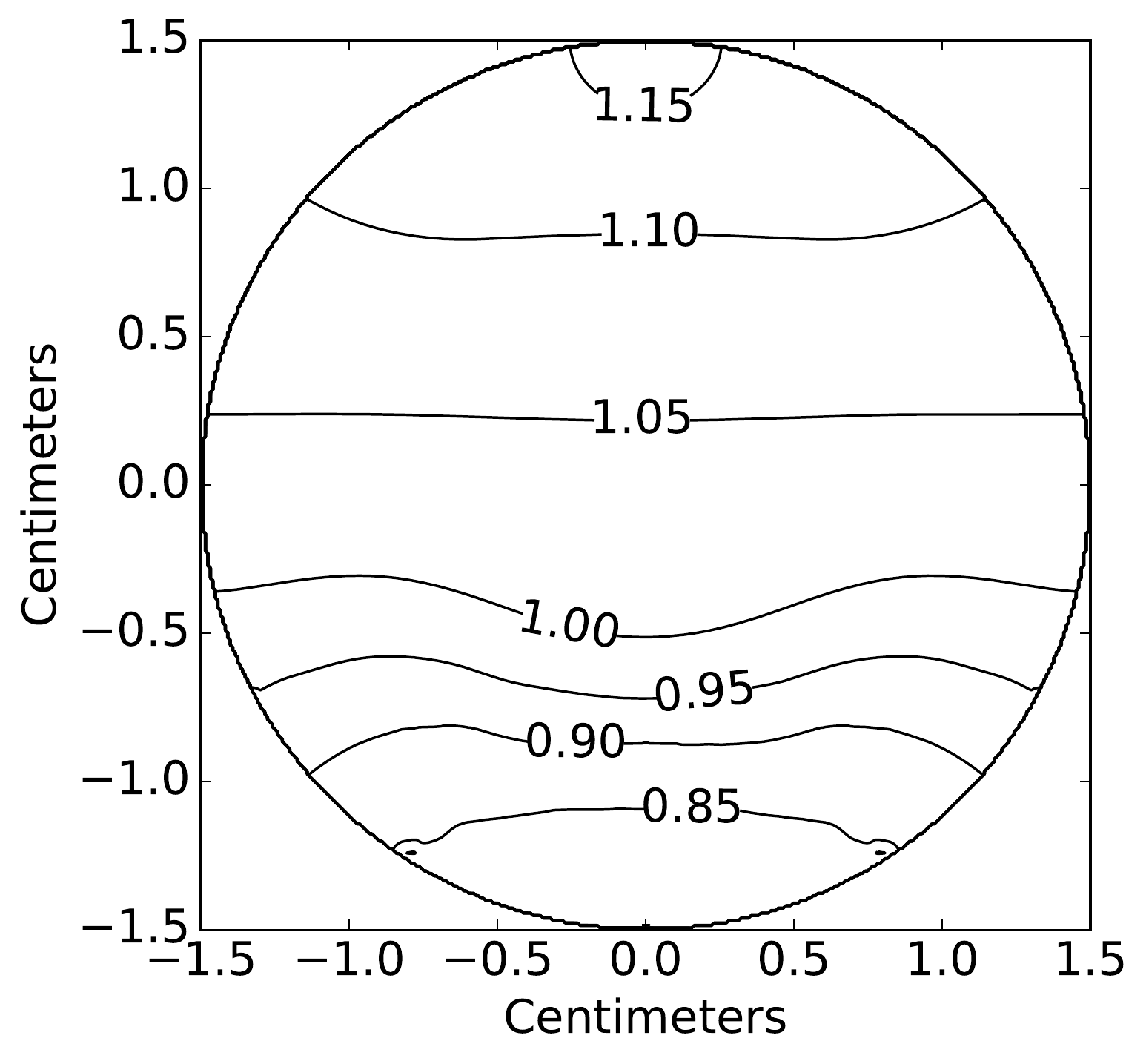}}
\subfloat[ART Reconstruction]{\label{fig:virtexpreconstruction}
  \includegraphics[width=0.49\columnwidth]{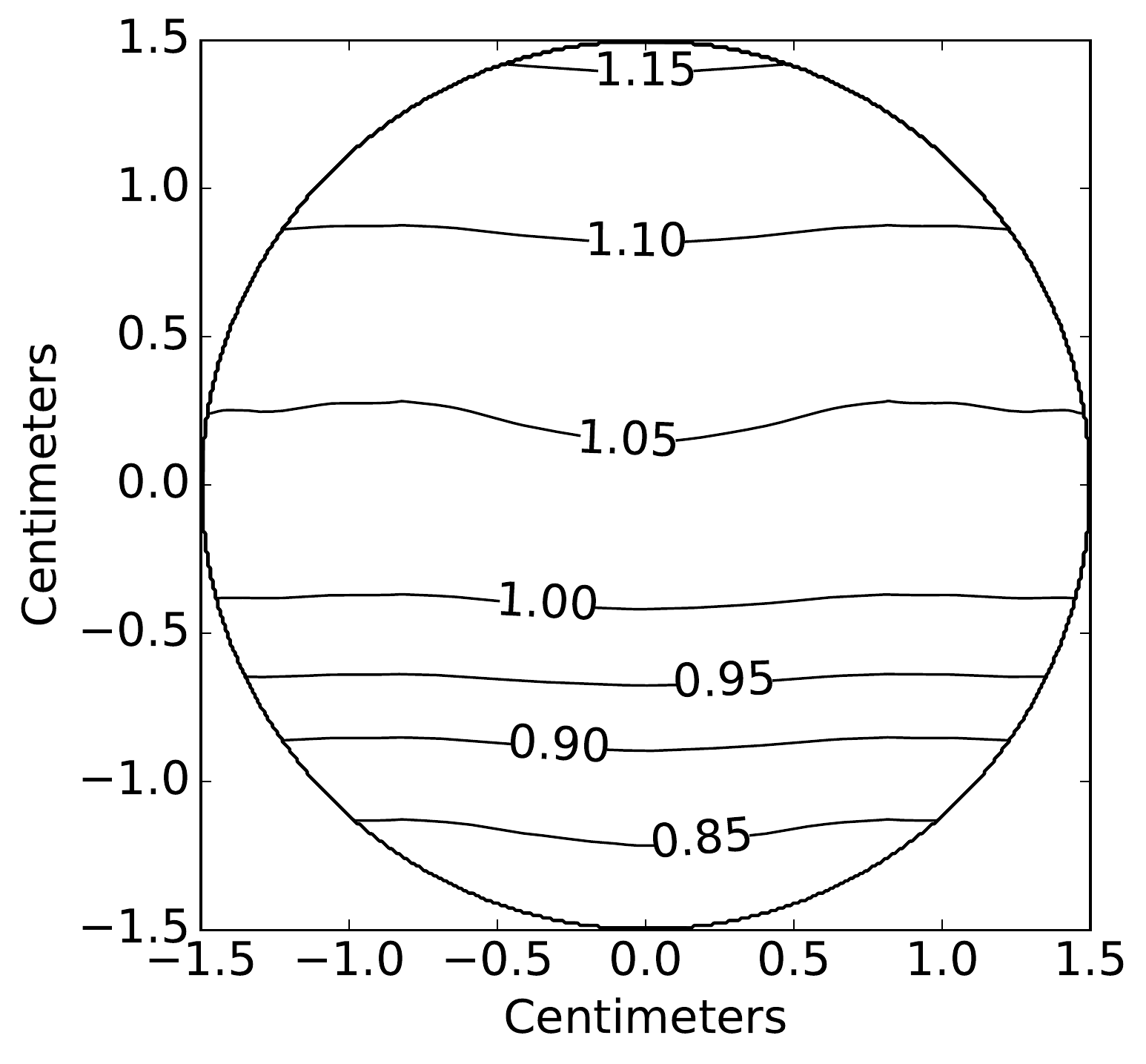}}
\caption{A virtual experiment where a representative density field is extracted from a ceramic powder compaction simulation and reconstructed from two orthogonal projections using multiplicative ART.
The \colorme{two-projection} ART reconstruction method adequately reconstructs the \colorme{general trends of the simulated density} field, although it does smooth along the projection directions.
Because the method is insensitive to the magnitude of the density, the density field of the simulated green body is normalized according to the average density of the slice.}
\label{fig:virtexp}
\end{figure}

A virtual experiment is now presented with the purpose of examining the efficacy of the two-projection algebraic reconstruction technique.
The examination is performed by comparing the reconstruction of a density field to a known solution produced by finite element software using a ceramic powder compaction model similar to that used by Stupkiewicz \textit{et al.}\cite{Stupkiewicz2014b}.
In the simulated experiment, a ceramic powder is compacted in a circular mold with a flat punch and a $10^\circ$ inclined base to form a green body in the shape of a truncated cylinder (such as the sample depicted in Figure \ref{fig:texture}).
The simulation was performed using Abaqus Standard and a user material routine of the above-mentioned model with 425 reduced-integration 3D hexehedral elements for a full 3D simulation of the compaction process.
A representative transverse slice of the simulated green body is used to generate two orthogonal projections with one of the projections being in line with the only plane of symmetry.
The results of the reconstruction, as well as the simulated green body density field, are presented in Figure \ref{fig:virtexp}.

For this virtual experiment, there is a consistent correlation between the density isosurfaces of the simulation and reconstruction but also a lack of curvature in the reconstruction that is present in the simulated density field.
Nevertheless, the reconstructed density field sufficiently resembles the aggregate simulated density field to the extent that the location, magnitude, and size of large-scale density variations can be identified \colorme{and compared with simulations to validate numerical models}.

When using this density evaluation method to compare a simulated density field to an experimental field, it is recommended that both the simulation and the experiment are to be reconstructed from projections and the resulting reconstructions compared.
In this way, the agreement between the simulation and experiment can be assessed in the same reconstruction space, subject to the same dissipation effects inherent to each reconstruction technique in order to yield two images that can be objectively compared.
This allows for the case that the simulation accurately predicts the experimental density field but that the reconstructed field does not exactly represent the actual solution.
As discussed in Section \ref{subsec:numberofprojections}, in only the most pathological instances will the reconstructed field exactly represent the actual field, such as in Figure \ref{fig:rog}.
While the set of density fields that satisfy the two projections is infinite, it is deemed unlikely that both a simulation and experiment would have the same projections with fundamentally different density fields.

\section{Green Body Reconstruction}

To complement the virtual experiment, the density analysis technique was used to analyze the truncated-cylinder green body depicted in Figure \ref{fig:texture}, with a diameter of $30.0\textrm{mm}$, a maximum \colorme{height} of $9.1\textrm{mm}$, and an upper-face inclination of $10^\circ$.
The green body was formed under a mean axial stress of $120\textrm{MPa}$, which is slightly more than the supplier-\colorme{recommended} $100\textrm{MPa}$ forming pressure to attain a green density of $2.4\textrm{g/cm\textsuperscript{3}}$.
The analysis took advantage of the symmetry of the green body along the diameter, thereby yielding two mirror-images of the sample.
These two sections of the green body were milled in perpendicular directions to give the minimum two projections to perform the reconstruction.

The sample was milled in transverse slices with a thickness of $1\textrm{mm}$.
Each transverse slice was partitioned into $1.5\textrm{mm}$-wide strips (one-half of the sample milled in one direction and the other half in the perpendicular direction) and then progressively milled the partitioned sections and weighed.
After measuring the projections of each transverse slice, the data can immediately be used to generate a reconstruction with a voxel size of $1.5\textrm{mm}\times1.5\textrm{mm}\times1.0\textrm{mm}$.

\begin{figure}
\centering
\includegraphics[width=0.9\columnwidth]{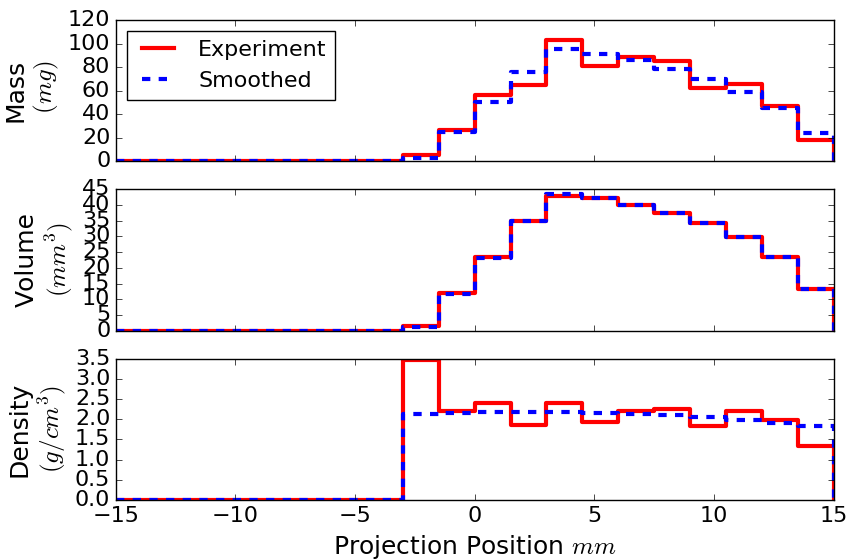}
\caption{A demonstration of the method to regularize the projections in the presence of experimental variation (such as flaking) during the milling process. The experimentally-measured mass can be converted to density by dividing by projection-element volume (found by using sample geometry and milling path). The density profile can then be smoothed by minimizing a mass-weighted root-mean-squared error function while requiring total mass to be unchanged.}
\label{fig:smoothexample}
\end{figure}

\begin{figure}
\centering
\subfloat[3D Render]{\label{fig:3ddensity}
  \includegraphics[width=0.9\columnwidth]{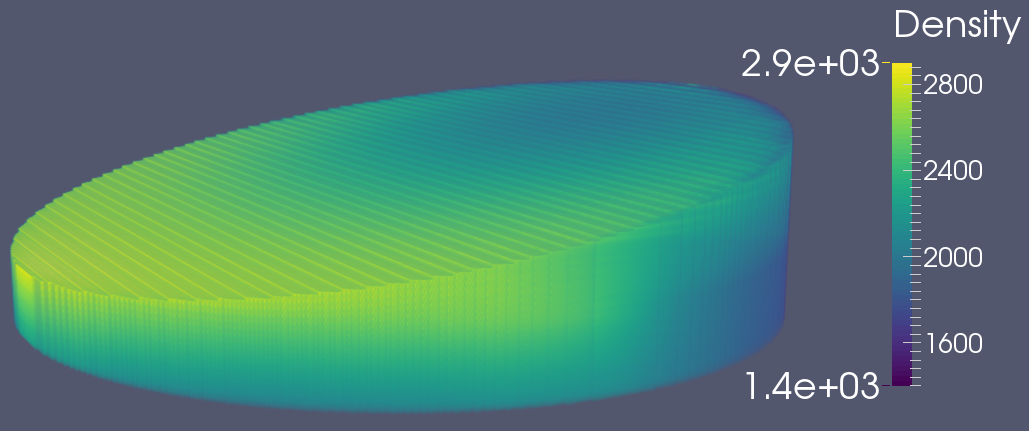}}
\\
\subfloat[Axial Cross-Section]{\label{fig:2ddensity}
  \includegraphics[width=0.9\columnwidth]{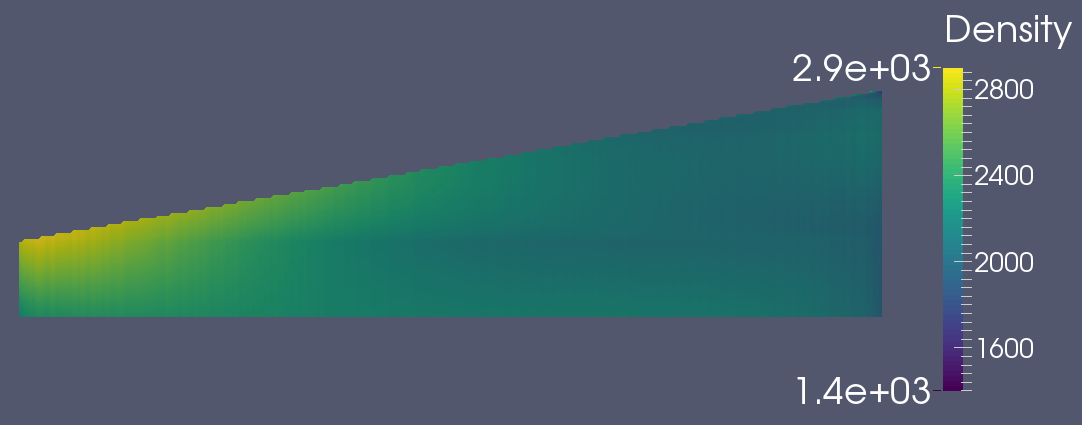}}
\\
\subfloat[Transverse Cross-Section]{\label{fig:2ddensitytrans}
  \includegraphics[width=0.9\columnwidth]{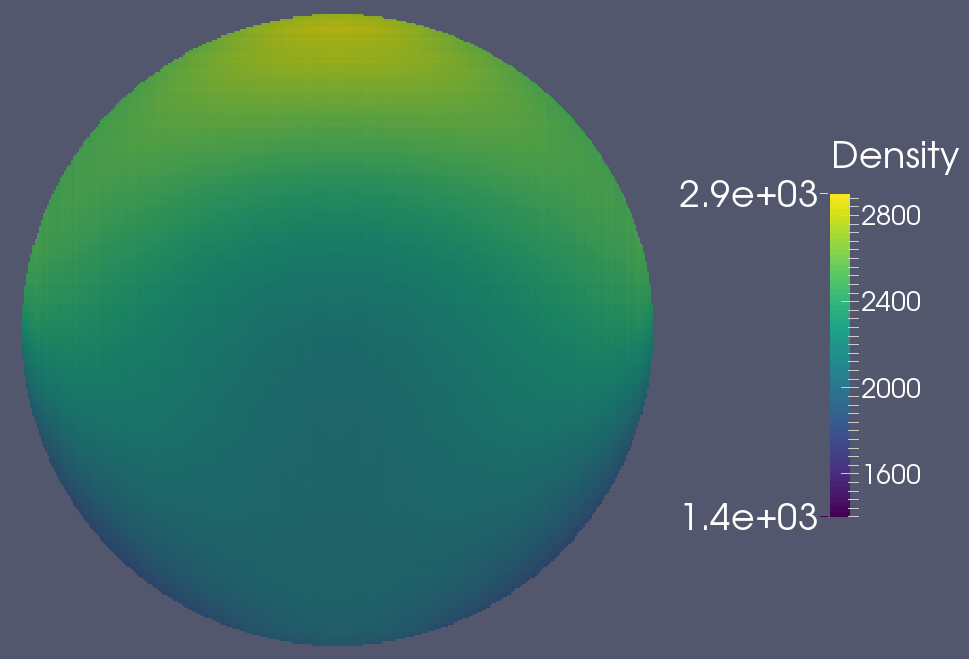}}
\caption{Visualized experimental data on a truncated cylindrical green body made of alumina powder (see Figure \ref{fig:texture} for geometry specifications). All densities are given in $\textrm{kg/m\textsuperscript{3}}$ with $1200\textrm{kg/m\textsuperscript{3}}$ as the uncompressed powder density and $2400\textrm{kg/m\textsuperscript{3}}$ as the green density at $100\textrm{MPa}$. The projections have been smoothed and then re-sampled from a strip width of $1.5\textrm{mm}$ to $0.1\textrm{mm}$.}
\label{fig:smootheddensity}
\end{figure}

During analysis, it was found that the milled strips that contained relatively \colorme{little} mass were particularly sensitive to the flaking of the sample during milling causing significant variation in the calculated density for that piece.
To overcome this, a smoothing step of the transverse slice projections was performed where the milled mass is transformed to density space and smoothed by minimizing a mass-weighted root-mean-squared error function subject to conservation of mass (see Figure \ref{fig:smoothexample}).
This gives more weight to the strips that had more mass and, therefore, a more accurate density value.
Applying this smoothing step to all the projections allows for resampling of the projection data from the smoothed projections to get a higher resolution reconstruction.
The high-resolution 3D reconstruction of the green body can be found in Figure \ref{fig:smootheddensity}.

While two projections are unlikely to exactly reproduce the density field, general predictions about the density field can still be readily inferred from the reconstruction.
From Figure \ref{fig:smootheddensity}, it can be seen that the bulk of the green body has a density that is approximately the reported green density for our alumina powder ($2.4\textrm{g/cm\textsuperscript{3}}$).
There is also an area of much higher density, approaching $2.9\textrm{g/cm\textsuperscript{3}}$, at the pinch point and that the other areas of higher density are more localized at the corners of the inclined surface leaving a relatively lower density in the center.

\section{Conclusion}
\label{sec:conclusion}

A simple, destructive method has been presented for determining internal density fields of ceramic green bodies, using only readily-available laboratory equipment.
The method has been demonstrated as being able to represent location, magnitude, and extent of large-scale density variations with sufficient accuracy \colorme{ for initial numerical model validation using the minimum two projections}.
Virtual experiments and experimental reconstructions have confirmed the utility of this method for determining and comparing density fields of green bodies and is now ready for use in research and industrial applications.

\section*{Acknowledgments}

M.S.S. gratefully acknowledges financial support from the European Union’s Seventh Framework Programme FP7/2007-2013/ under REA grant agreement number PITN-GA-2013-606878-CERMAT2.
A.P. gratefully acknowledges financial support from the Italian Ministry of Education, University and Research in the framework of the FIRB project 2010 Structural mechanics models for renewable energy applications. 
D.B. gratefully acknowledges financial support from ERC-2013-ADG-340561-INSTABILITIES.





\bibliographystyle{plain} 



\end{document}